\documentclass[superscriptaddress,twocolumn,secnumarabic,amssymb,amsmath,nobibnotes,aps,prd,showpacs,nofootinbib]{revtex4}

\pdfoutput=1
\usepackage{graphicx,subfigure,bm,color,psfrag,hyperref}
\usepackage{amsfonts}
\usepackage{lipsum}
\usepackage{caption}
\usepackage{mathtools}
\usepackage{verbatim}

\newcommand{\be}{\begin{equation}}
\newcommand{\ee}{\end{equation}}
\usepackage{changes}
\begin{document}

\title{Primordial gravitational waves in Horndeski gravity}

\author{Rafael C. Nunes}
\email{rafadcnunes@gmail.com}
\affiliation{Divis\~ao de Astrof\'isica, Instituto Nacional de Pesquisas Espaciais, Avenida dos Astronautas 1758, S\~ao Jos\'e dos Campos, 12227-010, SP, Brazil}

\author{M\'arcio E. S. Alves}
\email{marcio.alves@unesp.br}
\affiliation{Universidade Estadual Paulista (UNESP), Instituto de Ci\^encia e Tecnologia \\ S\~ao Jos\'e dos Campos, SP, 12247-004, Brazil}

\author{Jos\'e C. N. de Araujo}
\email{jcarlos.dearaujo@inpe.br}
\affiliation{Divis\~ao de Astrof\'isica, Instituto Nacional de Pesquisas Espaciais, Avenida dos Astronautas 1758, S\~ao Jos\'e dos Campos, 12227-010, SP, Brazil}

\begin{abstract}
We investigate the propagation of primordial gravitational waves within the context of the Horndeski theories, for this, we present a generalized transfer function quantifying the sub-horizon evolution of gravitational waves modes after they enter the horizon. We compare the theoretical prediction of the modified primordial gravitational waves spectral density with the aLIGO, Einstein telescope, LISA, gLISA and DECIGO sensitivity curves. Assuming reasonable and different values for the free parameters of the theory (in agreement with the event GW170817 and stability conditions of the theory), we note that the gravitational waves amplitude can vary significantly in comparison with general relativity. We find that in some cases the gravitational primordial spectrum can cross the sensitivity curves for DECIGO detector with the maximum frequency sensitivity to the theoretical predictions around 0.05 - 0.30 Hz. From our results, it is clear that the future generations of space based interferometers can bring new perspectives to probing modifications in general relativity.
\end{abstract}

\keywords{Modified gravity, Primordial gravitational waves}

\pacs{04.50.Kd; 04.30.−w}

\maketitle
\section{Introduction}

The LIGO collaboration reported the first direct detection of gravitational waves (GWs) through the GW150914 event \cite{ligo01}. Some time later other GWs events have been reported ~\cite{Gw02,Gw03,Gw04,Gw05,Gw06}.
Recently, the multimessenger astronomy arises with the detection of a binary neutron star merger by the LIGO and Virgo interferometers (GW170817 event \cite{Gw07}), and subsequently with the electromagnetic counterparts (GRB 170817A event \cite{Gw08}).
All these detections indicate a new era in modern astrophysics and cosmology, opening a new spectrum of possibilities to investigate
fundamental physics. More specifically,  in the cosmological context, the GW170817 event 
has imposed strong constraints on modified gravity/dark energy models \cite{GW_MG01,GW_MG02,GW_MG03,GW_MG04, GW_MG06, GW_MG07}.

An important source of GWs not detected until the present time are GWs of cosmological origin, i.e., the primordial gravitational waves (PGWs). The future detection of such waves by space-borne interferometers, or by the measurements of the B-mode of polarization of the cosmic microwave background (CMB) radiation, will bring unique information about the physics of the early Universe. This is because the PGWs spectrum is sensitive to the evolution of the Universe in the inflationary epoch in which the scale factor grows exponentially, while the Hubble horizon is kept constant. In this scenario, the initial quantum tensor modes are inside the Hubble volume, and become effectively classical as the Universe expands and they leave the horizon. This quantum-to-classical transition provides the metric perturbation, of quantum origin, equivalent to a stochastic variable in the Hubble crossing. The perturbations re-enter progressively the Hubble horizon during the evolution of the Universe, leading to a GW signal which is, therefore, intrinsically stochastic (see, e.g. \cite{Caprini}, for a review).

Although they have not yet been detected, an upper bound of PGWs in a specific scale can be currently quantified through the tensor-to-scalar ratio $r$ parameter from the CMB data. The current borders 
are $r < 0.10$, by Planck team within the minimum $\Lambda$CDM model at 95\% confidence level by combining the spectra of temperature fluctuations, low-polarization, and lensing \cite{Akrami:2018odb}. When combined in a joint analysis with BICEP/Keck CMB polarization experiments, we have tighter borders, namely, $r < 0.06$ \cite{bicep2}. However it is expected that the future generations of space interferometers could detect the PGWs, or even put strong bounds in their amplitudes. Contrary to the ground-based LIGO interferometer, which has a sensitivity frequency band ranging from 10 Hz to 1 kHz, space-based GWs detectors are able to achieve lower frequencies for which the inflationary PGWs are expected to have higher amplitudes. The most notable example of a space interferometer, which has been under study for several years is the LISA mission, aiming to detect GWs in the $10^{-4} - 1$ Hz band \cite{Amaro2017}. On the other hand, the proposed space mission DECIGO intends to detect GWs in a frequency band located between LISA and LIGO (0.1 Hz to 10 Hz) \cite{Seto2001}. In a similar frequency band, a geosynchronous version of the LISA detector (gLISA) has also been proposed in order to operate simultaneously with LISA \cite{Tinto2016}. The frequencies of the order of nanohertz, on the other hand, can be achieved only by the pulsar timing technique, specifically by using arrays of millisecond pulsars. At this time, efforts are underway in order to improve the sensitivity in this band \cite{Verbiest2016}. 

In practice, the spectrum of PGWs is not only determined by the evolution of the background cosmology, but it can be significantly affected by modifications in the General Relativity (GR) theory \cite{MG_GW01,MG_GW02,MG_GW03,MG_GW04,MG_GW05,MG_GW06, MG_GW07,MG_GW08, MG_GW09, MG_GW10, MG_GW11, MG_GW13, MG_GW14, MG_GW15}, or on early physical aspects \cite{inflation_GW01,inflation_GW02,inflation_GW03, inflation_GW04, inflation_GW05}, such as an inflationary phase.
In this work, our aim is to investigate the propagation of the PGWs in the context of the Horndeski gravity. In \cite{Deffayet} Deffayet et al. derived the action of the most general scalar-tensor theories with second-order equations of motion after the generalizations of covariant Galileons. In \cite{Kobayashi} it is shown that the corresponding action is equivalent to that derived by Horndeski in 1974 \cite{Horndeski}. 
Because it is a general theory of gravitation, once different modified gravity theories predict different cosmic evolution, it is possible to distinguish between scenarios in Horndeski theories from observations \cite{Horndeski_constraints_01,Horndeski_constraints_02,Horndeski_constraints_03,Horndeski_constraints_04,Horndeski_constraints_05, Horndeski_constraints_06, Horndeski_constraints_07, Horndeski_constraints_08,Horndeski_constraints_09,Horndeski_constraints_10}. 

As the main result of this work, we present a generalized transfer function quantifying the propagation of the PGWs within Horndeski theories and we evaluate the present theoretical spectrum and compare it with the sensitivity curves of different GW experiments, such as aLIGO \cite{aLIGO}, DECIGO \cite{DECIGO}, ET \cite{ET} and LISA \cite{eLISA}.  In \cite{MG_GW04} the authors also present how the modified GW propagation can affect the transfer function. Here, we show a more general transfer function, which is compatible with \cite{MG_GW04} if the time delay factor is set to zero. Moreover, we find that the spectra can significantly differ from that predicted by GR and, therefore, can in the future be probed observationally.

The manuscript is organized as follows: In Section \ref{sec-model}, we introduce a method to calculate the GW energy spectrum in the context of the Horndeski gravity. In Section \ref{results2}, the prediction for the present spectrum of PGWs is evaluated and compared with the sensitivity curves of different GW detectors. Finally, in Section \ref{Conclusions} we summarize our findings and conclude with our final remarks. As usual, a sub-index zero attached to any physical quantity refers to its value at the present cosmic time. Also, prime and dot denote the derivatives with respect to the conformal time and cosmic time, respectively.

\section{Primordial gravitational waves in the context of the Horndeski gravity}
\label{sec-model}

The Horndeski theories of gravity \cite{Horndeski, Deffayet} are the most general Lorentz invariant scalar-tensor theories with second-order equations of motion. The Horndeski action reads

\begin{equation}
\label{acao_geral}
 S =  \int d^4 x \sqrt{-g} \Big[ \sum_{i=2}^{5} \frac{1}{8 \pi G} \mathcal{L}_i + \mathcal{L}_m \Big],
\end{equation}

\begin{equation}
 \mathcal{L}_2 = G_2(\phi, X),
\end{equation}

\begin{equation}
 \mathcal{L}_3 = -G_3(\phi, X) \Box \phi,
\end{equation}

\begin{equation}
 \mathcal{L}_4 = -G_4(\phi, X)R + G_{4X} [( \Box \phi)^2 - \phi_{;\mu \nu} \phi^{;\mu \nu}],
\end{equation}

\begin{align}
\mathcal{L}_5 = -G_5(\phi, X)G_{\mu \nu}\phi^{;\mu \nu} -
\frac{1}{6}G_{5X}[(\Box \phi)^3  + \\
2 \phi_{;\mu \nu}  \phi^{;\mu \sigma} \phi^{;\nu}_{;\sigma}
 - 3 \phi_{;\mu \nu} \phi^{;\mu \nu} \Box \phi],
\end{align}
where the functions $G_i$ ($i$ runs over 2, 3, 4, 5) depend on $\phi$ and $X = -1/2 \nabla^\nu \phi \nabla_\nu \phi $, with $G_{i X} = \partial G_i/\partial X$. For $G_2 = \Lambda$, $G_4 = M^2_p/2$ and $G_3 = G_5  = 0$, we recover GR with a cosmological constant. For a general discussion on the model varieties for different $G_i$ choices see \cite{Tsujikawa}.

In the present work, we are particularly interested in the evolution of PGW through an expanding Universe. The  evolution of linear, transverse-traceless perturbations for the tensor modes due to modifications in the gravity theory is generally described by the following equation \cite{Saltas}
\begin{equation}
\label{h_general}
\ddot{h}_{ij} + (3 + \nu) H \dot{h}_{ij} + ( c_T^2 k^2/a^2 + \mu^2)h_{ij} = \Gamma \gamma_{ij}, 
\end{equation}
where $h_{ij}$ is the metric tensor perturbation. The four time dependent parameters are:  $c_T$ is the GW propagation speed, $\mu$ is the effective graviton mass, $\nu$ is related to the running of the effective Planck mass, and $\Gamma$ denotes extra
sources generating GWs.

In the context of the Horndeski gravity, the above equation reads
\begin{equation}
\label{new_h}
\ddot{h}_{ij} + (3 + \alpha_M) H \dot{h}_{ij} + (1 + \alpha_T) \frac{k^2}{a^2} h_{ij} = 0, 
\end{equation}
where we have identified $\nu = \alpha_M$, $c_T^2 = 1 + \alpha_T$, $\mu = 0$ and $\Gamma = 0$, where $\alpha_M$ and $\alpha_T$ are two dimensionless functions given by
\begin{equation}
\label{alphaM}
\alpha_M = \frac{1}{H M^2_{*}} \frac{dM^2_{*}}{dt},
\end{equation}

\begin{equation}
\label{alphaT}
\alpha_T = \frac{2 X (2G_{4X} - 2G_{5 \phi} - (\ddot{\phi} - \dot{\phi} H)G_{5X})}{M^2_{*} },
\end{equation}
and $M_{*}$ is the effective Planck mass
\begin{equation}
M^2_{*} = 2(G_4 - 2XG_{4X} + XG_{5 \phi} - \dot{\phi} H X G_{5X}). 
\end{equation}

The running of the Planck mass, $\alpha_M$, enters as a friction term and it is responsible for modifying the amplitude of the tensor modes acting as a damping term. But it is also related to the strength of gravity. On the other hand, the tensor speed excess, $\alpha_T$, modifies the propagation speed of the GWs quantifying a modification on the GW phase. As can be seen in the above equations, the functions $\alpha_M$ and $\alpha_T$  depend on the parameters of the theory and on the cosmological dynamics of the scalar field. 

Following the methodology presented in \cite{MG_GW02}, we can describe the sub horizon evolution of GWs in a modified gravity theory as
\begin{equation}\label{hDT}
 h = e^{-\mathcal{D}} e^{- i k \Delta T} h_{GR},
\end{equation}
where
\begin{equation}\label{DD}
\mathcal{D} = \frac{1}{2} \int^{\tau} \alpha_M \mathcal{H} d\tau',
\end{equation}

\begin{equation}\label{DTD}
\Delta T = \int^{\tau} (1 - \sqrt{1 + \alpha_T}) d\tau',
\end{equation}
where $\mathcal{D}$ and $\Delta T$ correspond to the amplitude damping and
additional time delay of the GWs, respectively. Consequences of the Horndeski theory at cosmological scale were recently investigated in \cite{MG_GW03}. Here and through the text, $\tau$ represents conformal time.

In \cite{MG_GW02} the above equations were obtained in the WKB approximation for which the GW wavelength is much smaller than the cosmological horizon. In the case of PGWs, the modes leave the horizon in the inflationary period, and the modes with physical frequency $f \gtrsim 10^{-15}$ Hz reenter the horizon in the radiation era (see, e.g., \cite{TF01}). Therefore, in the milihertz frequency band in which the future LISA detector will operate, or for higher frequencies, it is reasonable to use this WKB solution to evaluate the evolution of GWs in the subsequent stages after inflation. This is because the change in the amplitude is a cumulative effect throughout the propagation of the GWs, and for these frequencies, its wavelength is much shorter than the horizon in the most part of the time of evolution. The initial conditions of such an evolution are obtained at the end of inflation \cite{TF01}.


The GW is usually characterized by its amplitude $h(k, \tau)$ or by its energy spectrum $\Omega_{GW} (k, \tau)$. Here, we are particularly interested in the GW spectrum, which in the standard context of GR is given by (see \cite{Caprini} and reference therein)

\begin{equation}
\label{OmegaGW_RG}
\Omega_{GW} (k, \tau) = \frac{1}{12 H^2 a^2} [T'(k, \tau)]^2 P_t(k),
\end{equation}
where $T(k, \tau)$ is the transfer function that describes the sub-horizon evolution of GW modes after the modes are deep inside the horizon. It is worth mentioning that the methodology for computing the transfer function has been widely discussed in the literature \cite{TF01,TF02, TF03}. The quantity $P_t(k)$ is the amplitude spectrum of GWs at the end of the inflationary period. Throughout our calculations, let us adopt
\begin{equation}\label{amplitude from inflation}
 P_t(k) = \frac{k^3}{2 \pi^2}(|h^{+}_k|^2 + |h^{\times}_k|^2 ) = A_t \Big(\frac{k}{k_{*}} \Big)^{n_t},
\end{equation}
where $A_t$ is the tensor amplitude at the reference scale $k_{*}$, and $n_t$ is the tensor spectral index. Here, $h^{+,\times}$ denotes the amplitude of the two polarization states $(+, \times)$ of GWs.

In what follows we are interested in generalizing Eq. (\ref{OmegaGW_RG}) in order to introduce the effects of the tensor propagation modes due to the modifications induced by Horndeski gravity given by Eq. (\ref{new_h}). By definition, we have that the transfer function is given by
\begin{equation}
T(k, \tau) =  \frac{h_k(\tau)}{h_k(\tau_i)},
\end{equation}
where $h_k(\tau_i)$ is the primordial GW mode that left the horizon during inflation.

Given the general formulation of GW propagation within the Horndeski scenario, we can write a new transfer function as
\begin{equation}
\begin{aligned}
\label{Tk_MG}
& T(k, \tau)_{MG}  =  \Big \{ \Big[ \exp \Big( - \frac{1}{2} \int^{\tau} \alpha_M \mathcal{H} d\tau' \Big) \Big] \\
& \times \Big[ \exp \Big( i k \int^{\tau} (\sqrt{1 + \alpha_T} - 1) d\tau' \Big) \Big] \Big \} T(k, \tau)_{GR},
\end{aligned}
\end{equation}
where $T(k, \tau)_{GR}$ is the standard transfer function of GR. The index ${\rm MG}$ in the above equation means modified gravity (in the present case, the Horndeski gravity). As expected, for $\alpha_M = \alpha_T = 0$, we recover GR. Substituting Eq. (\ref{Tk_MG}) into Eq. (\ref{OmegaGW_RG}), we can quantify the effects of the Horndeski gravity in terms of the functions $\alpha_i$ on the transfer function, and consequently on a new and generalized primordial energy spectrum $\Omega_{GW} (k, \tau)$. In what follows, in all the results to be presented in this work, to calculate $T(k, \tau)_{GR}$, we use the methodology presented in \cite{TF01}.

It is usual to choose phenomenologically motivated functional forms for the functions $\alpha_i$ (see, e.g., \cite{Bellini, alphai_01,alphai_02,alphai_03}). Typically, their evolution are tied to the scale factor $a(t)$ or to the dark energy density $\Omega_{de}(a)$ raised to some power $n$. In the present work we will adopt the following parametrization
\begin{equation}\label{parametrization}
\alpha_i =\alpha_{i0} a^n,
\end{equation}
where the label $i$ runs over the set of functions $M$ and $T$. Such a parametrization has been frequently considered in the literature, and it was recently suggested that this form encompasses the effects of the different modified gravity theories (see, e.g., \cite{Gleyzes2017}). Hence, this form is particularly suitable for comparing those theories with cosmological observations and, therefore, it is also useful for our present purposes.

On the other hand, the event GW170817 from a binary neutron star merger together with the
electromagnetic counterpart showed that the speed of GW, $c_T$, is very close to that of
light for $z < 0.01$, that is, $ |c_T/c - 1| < 10^{-15}$ \cite{Gw08}. Thus, as we are interested in calculating the spectrum at the present time, let us assume from now on that $\alpha_{T0}~=~0$ in Eq. (\ref{Tk_MG}), in full agreement with the GW170817 observation. Therefore, the correction on the time delay factor, which induces a phase shift on the transfer function, will not be taken into account.

An important point within Horndeski gravity are the stability conditions of the theory. Appropriate values of free parameters linked to the $\alpha_i$'s functions must be taken in order 
to have a stable theory throughout the evolution of the universe (see \cite{alphai_02} and reference therein). Once here the only parameterization that will model our results is $\alpha_M(a)$, that is, the amplitude damping correction on the standard prediction, let us only discuss the stability values on $\alpha_M(a)$ function. Following \cite{alphai_02}, adopting, $\alpha_M =\alpha_{M0} a^n$, we have that the stability conditions can be summarized as follow
\\

1. $n > 3/2$: Stable for $\alpha_{M0} < 0$.
\\

2. $0 < n < 3\Omega_{m0}/2$: Stable for $\alpha_{M0} > 0$,
\\
\\
where $\Omega_{m0}$ is the dimensionless matter density.
\\

Under these considerations, we can note from Eq. (\ref{Tk_MG}) that the changes in the GW spectrum will be  an increase in the amplitude for the stability conditions 1, and a decrease when considering the conditions 2. In what follows, let us only assume values of the pair ($n$, $\alpha_{M0}$) within of this range of values.

\begin{figure*}
\includegraphics[width=3.4in,height=2.5in]{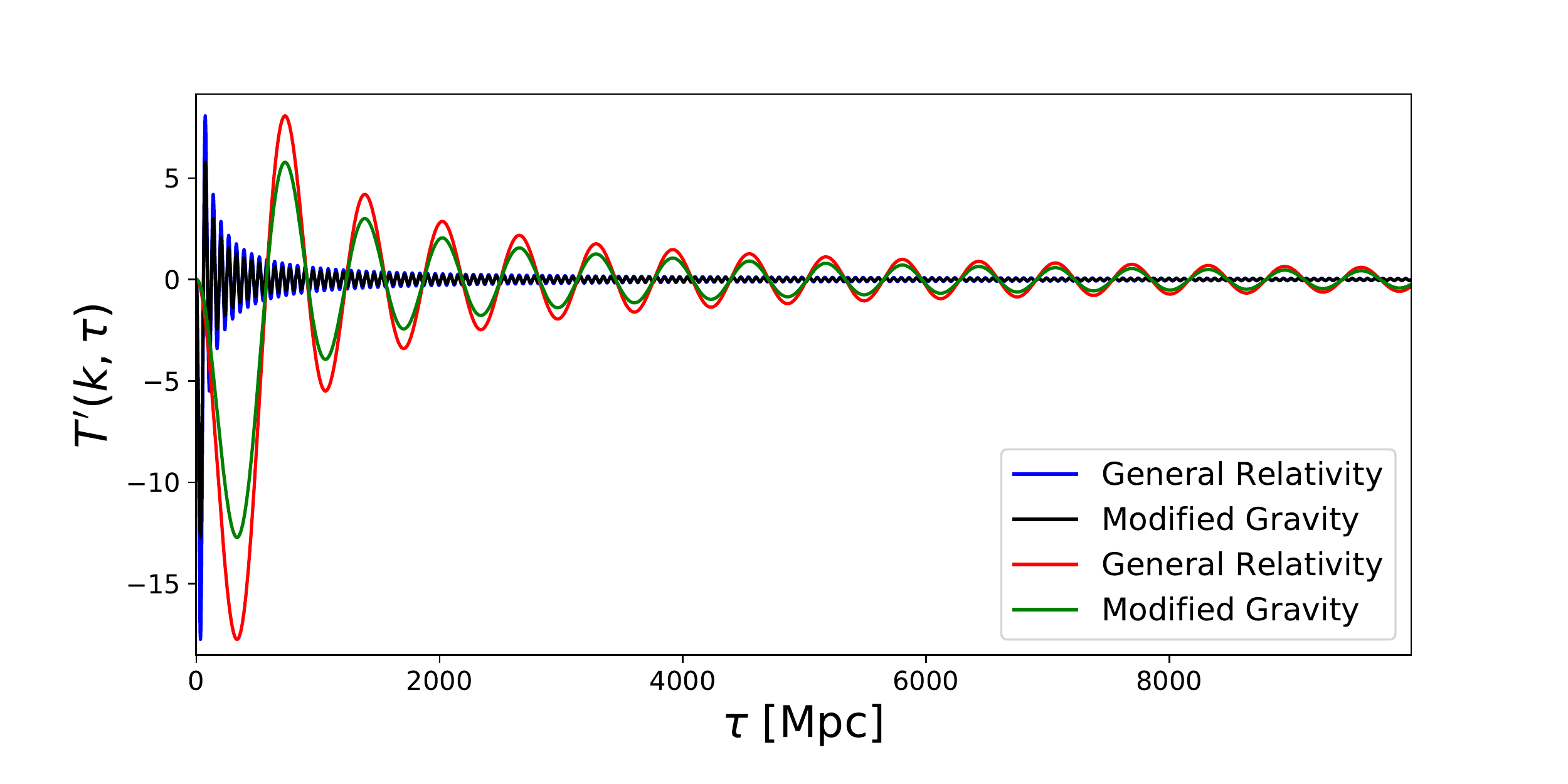} \quad 
\includegraphics[width=3.4in,height=2.5in]{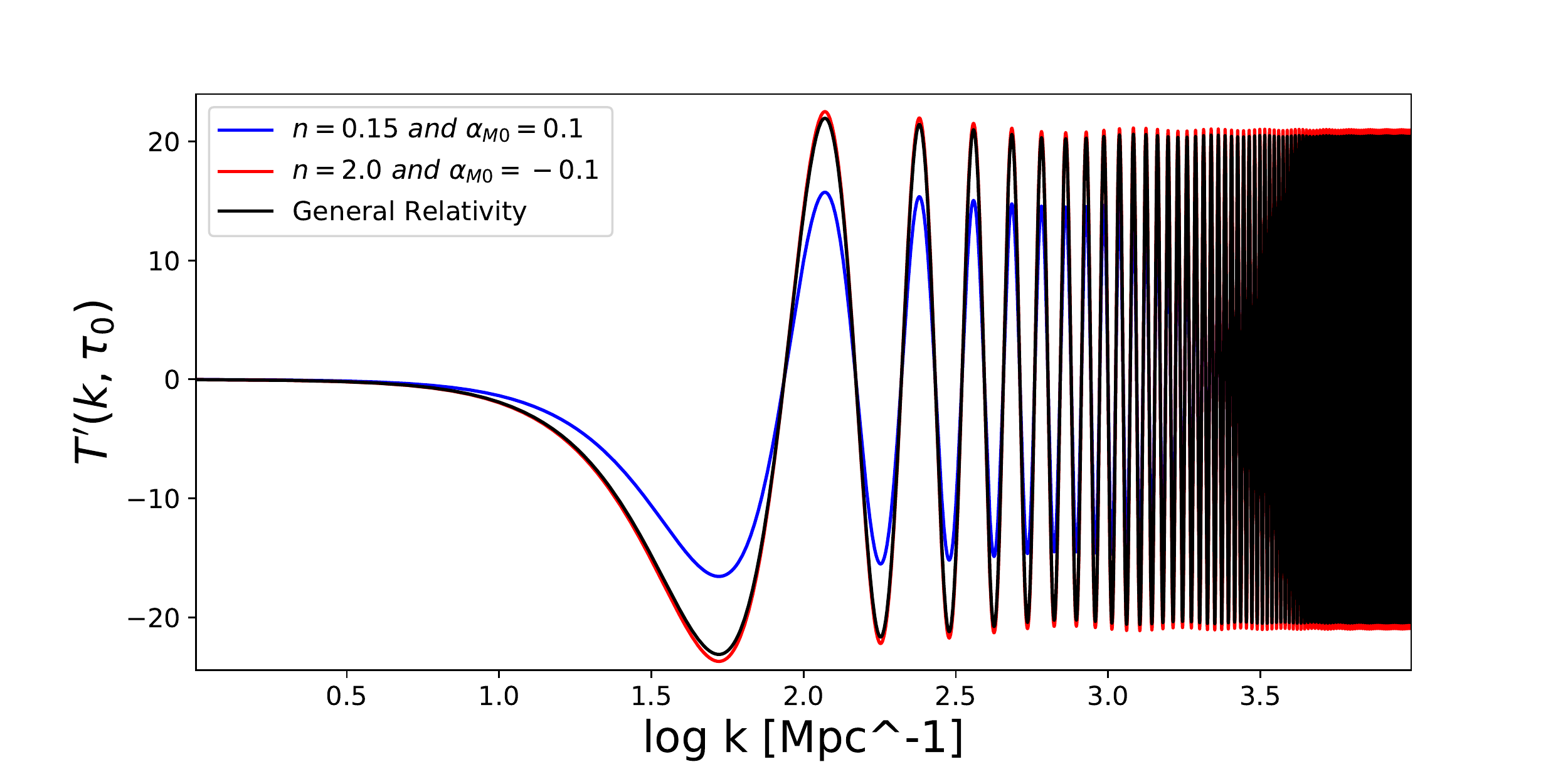}
\caption{Left panel: Evolution of the derivative of the transfer function, $T'(k, \tau) $, as a function of the conformal time for two different scales $k = 0.1$ ${\rm Mpc}^{-1}  $ (blue and black lines) and $k = 0.01$ ${\rm Mpc}^{-1} $ (red and green lines). The blue and red lines correspond to the standard evolution in GR, while the black and green lines show the modified behavior due to the Horndeski gravity with $n = \Omega_{m0}/2$ and $\alpha_{M0} = 0.1$. Rigth panel: Derivative of the transfer function evaluated today as a function of $k$.}
\label{TF}
\end{figure*}

\section{PGWs evolution and Spectrum}
\label{results2}
\vspace{5mm}

In the previous section, we saw that the modification introduced in the PGW spectrum by the Horndeski gravity, with respect to GR, is encapsulated in the transfer function given by Eq. (\ref{Tk_MG}). Now, in order to analyze the effects in the evolution of $T'(k, \tau)$ exclusively due to the modifications introduced by the gravity theory, let us consider that the background cosmology, for both theories, starts with the same inflationary era, followed by the subsequent usual radiation and matter eras. All the information regarding inflation comes only from the parametrized inflationary spectrum given by Eq. (\ref{amplitude from inflation}). The derivative of the transfer function $T'(k, \tau)$ is responsible for the further processing of such a spectrum during the expansion of the Universe, until the present time. 

Therefore, with the parametrization (\ref{parametrization}) and with the above considerations, the transfer function can be obtained in a straightforward way. In the left panel of Figure \ref{TF}, the evolution of $T'(k, \tau)$ as a function of the conformal time is shown, in comparison with the standard behavior obtained from GR. In this figure, two different scales are considered, namely, $k = 0.1$ ${\rm Mpc}^{-1}  $ and $k = 0.01$ ${\rm Mpc}^{-1} $. 
As already mentioned, the main effect is in the amplitude of the GWs, 
once modifications on the phase are not assumed, i.e, $\alpha_{T0} = 0$. If $\alpha_{M0}$ is positive, the higher is its value, the smaller is the amplitude of GWs, while $n$ is kept fixed and positive. Otherwise, assuming $\alpha_{M0} < 0$ and $n$ fixed (and positive), the GW amplitude increases.
Since we are particularly interested in evaluating the PGW spectrum at the present time, we also evaluate the present value of $T'(k, \tau)$ as a function of the wave number $k$. The result is shown in the right panel of Figure \ref{TF}. The corresponding modifications in the transfer function induced by the Horndeski gravity leave an imprint on $\Omega_{GW}$ resulting in a final spectrum that deviates from GR.



In order to compute the present energy density spectrum we need to consider Eq. (\ref{OmegaGW_RG}) evaluated at the present time $\tau_0$. In what follows, we will also assume that $n_t = - 0.01$ and $A_t = 10^{-10}$ in Eq. (\ref{amplitude from inflation}), in agreement with the last results of the Planck team \cite{Akrami:2018odb}. Moreover, a stochastic GW background is often characterized also by its spectral density $S_h(f)$\footnote{In order to make connection with observations, it is necessary to evaluate the GW background today in terms of the present-day physical frequency $f = k/2 \pi a_0$. The spectral density  $S_h(f)$ is given in $ {\rm Hz^{-1}}$.}. This quantity is better suited for a direct comparison with a GW detector. The relation between $S_h(f)$ and $ \Omega_{GW}(f)$ is as follows \cite{Caprini}

\begin{equation}
\label{GW_th}
\Omega_{GW}(f) = \frac{4 \pi^2}{3 H_0^2} f^3 S_h(f).
\end{equation}

\begin{figure*}
\includegraphics[width=3.5in,height=2.5in]{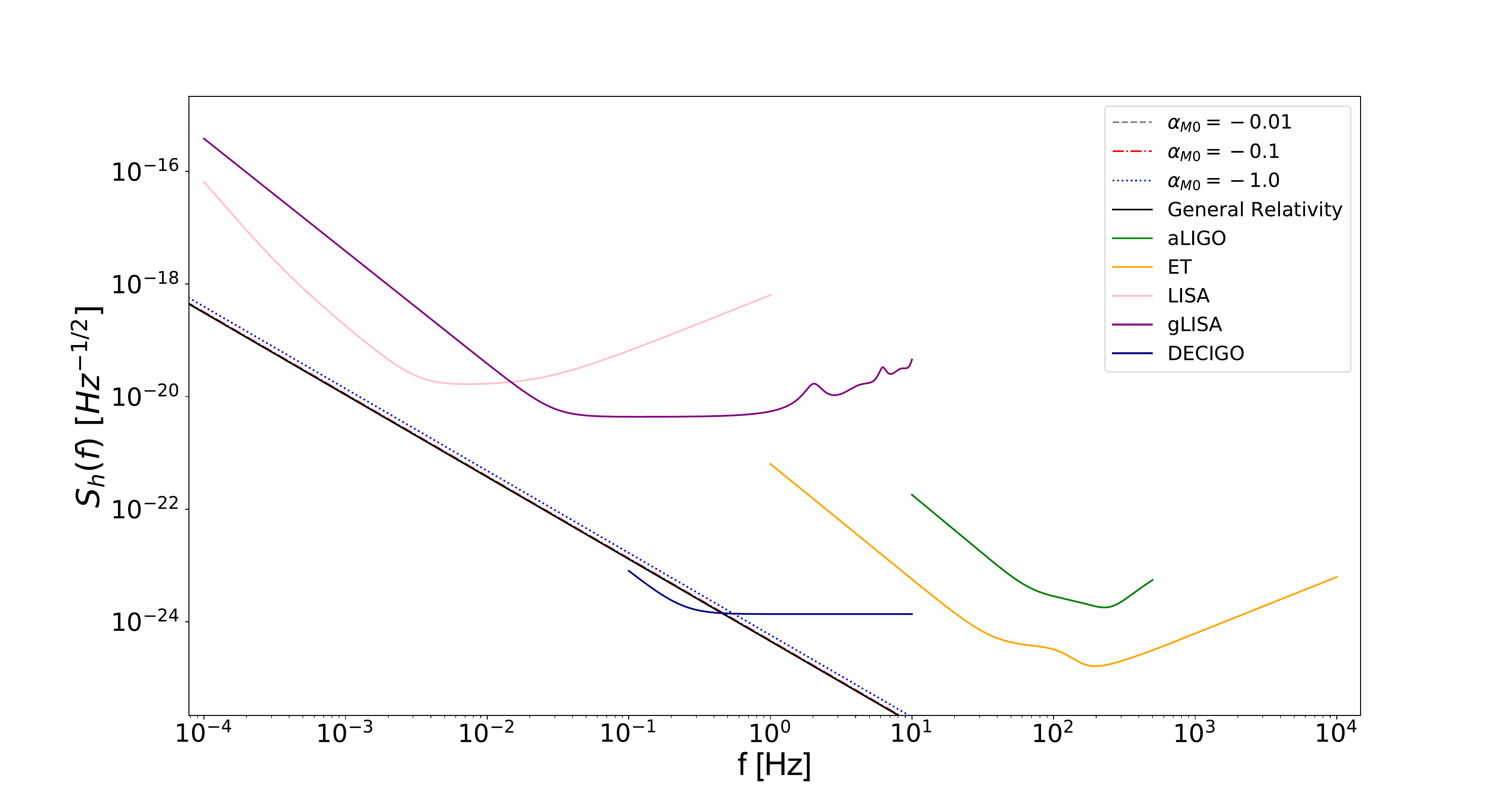} 
\includegraphics[width=3.5in,height=2.5in]{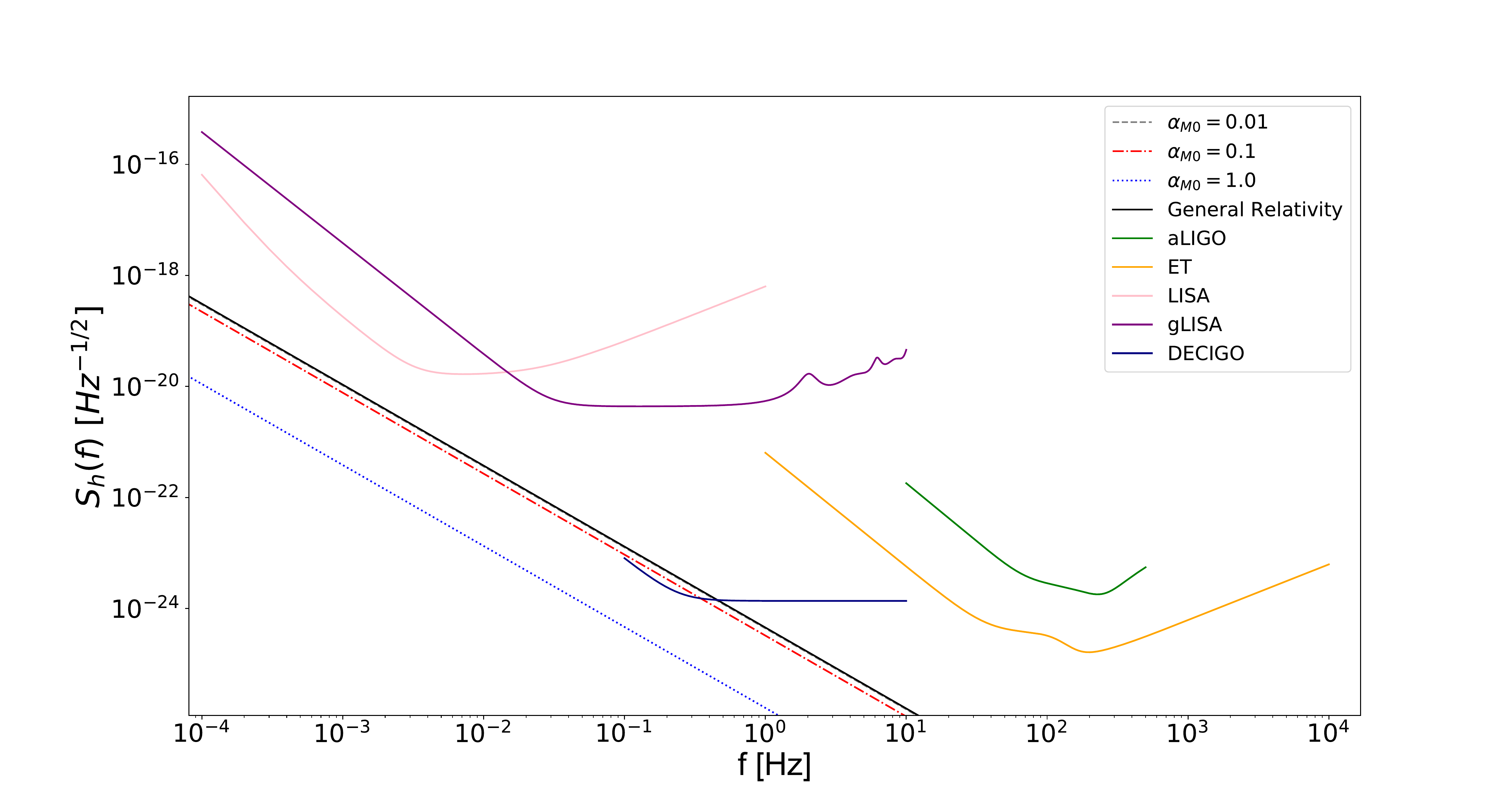}
\caption{Left panel: Theoretical prediction of the GW spectral density for some values of $\alpha_{M0} < 0$ with $n =2$ in all cases. Right panel: Same as the left panel, but for some values of $\alpha_{M0} > 0$ and $n = \Omega_{m0}/2$ fixed, where we take $\Omega_{m0} = 0.30$. 
All values are according to the stability condition of the theory. 
The predicted sensitivity curves for some ground and space based GWs detectors are also shown. }
\label{Sh1}
\end{figure*}

\begin{figure*}
\includegraphics[width=3.5in,height=2.5in]{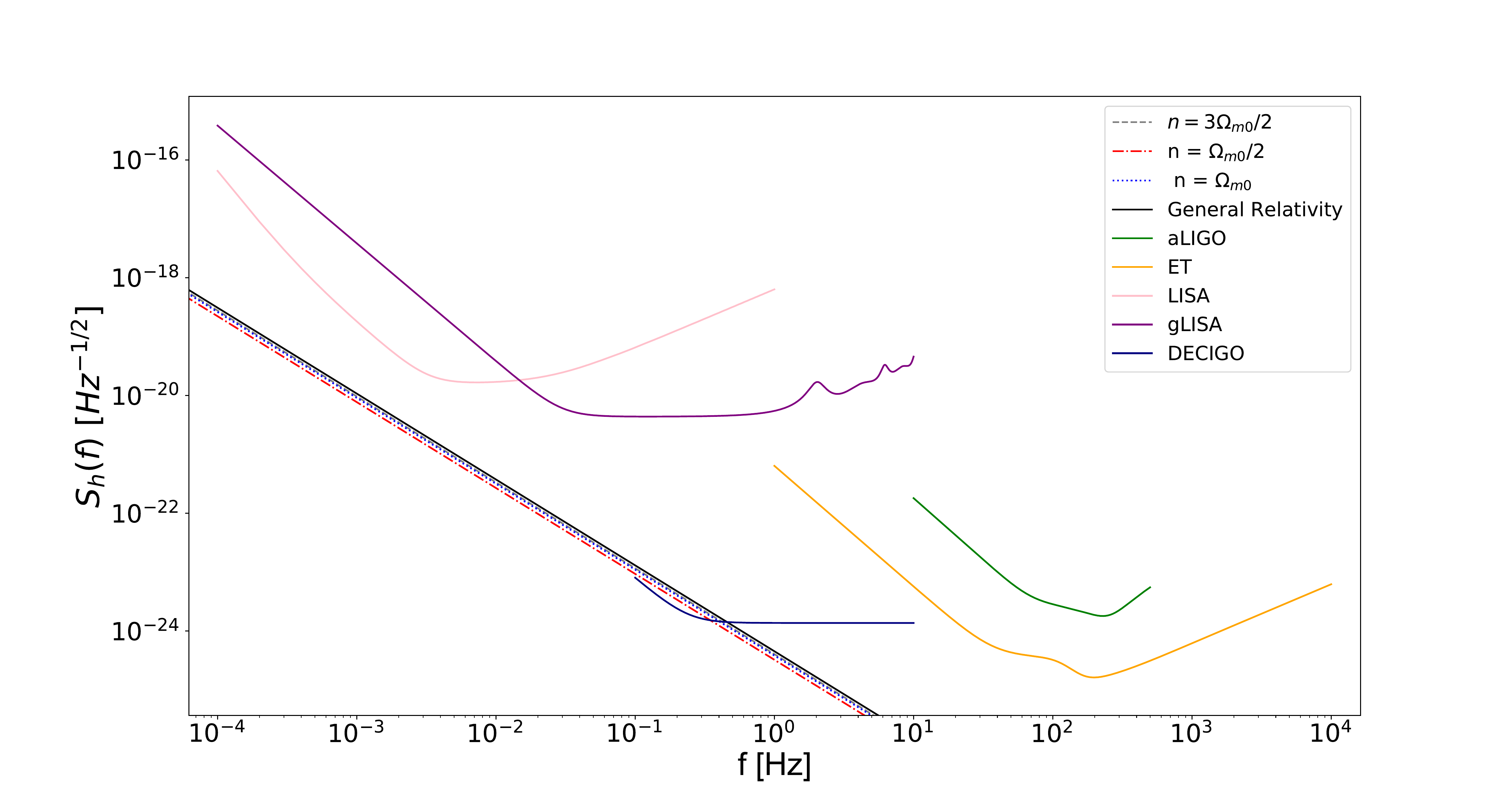}
\includegraphics[width=3.5in,height=2.5in]{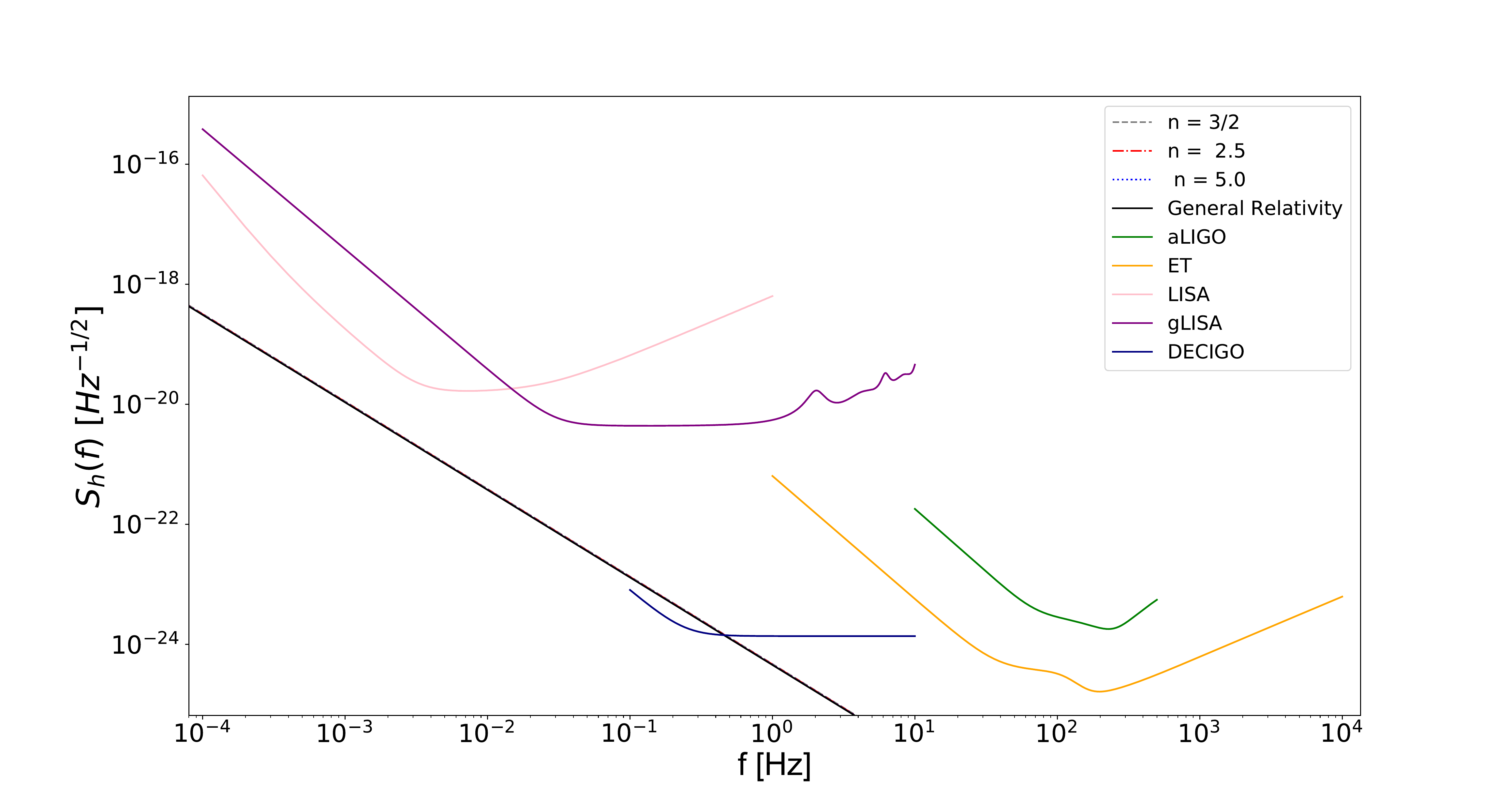}
\caption{Left panel: Theoretical prediction of the GW spectral density for some values of $n$ with $\alpha_{M0} = 0.1$ in all cases. Rigth panel: Same as the left panel, but for $\alpha_{M0} = -0.1$. In both panels we take $\Omega_{m0} = 0.30$. All values are according to the stability condition of the theory. The predicted sensitivity curves for some ground and space based GWs detectors are also shown. }
\label{Sh2}
\end{figure*}

\begin{figure}
\includegraphics[width=3.25in,height=2.5in]{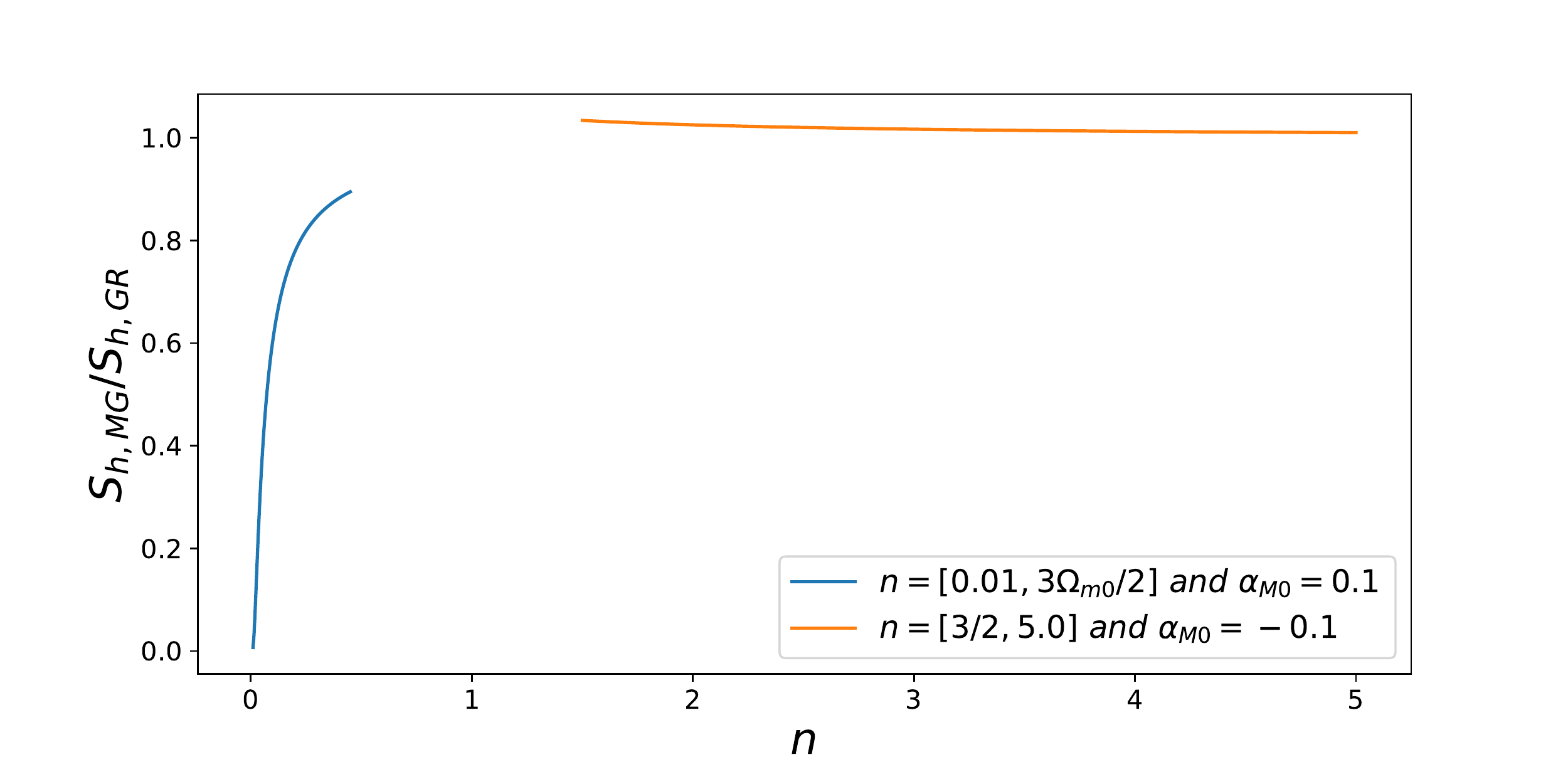}
\caption{Ratio of the GW spectral densities, $S_{h, MG}/S_{h, GR}$, as a function of $n$ keeping $\alpha_{M0}$ fixed, according to the predicted values within the stability conditions of the theory.}
\label{DeltaSh}
\end{figure}


In computing the PGW spectrum, we have considered the specific frequency bands of interest that cover the sensitivity curves of some ground and space based interferometers. For the ground interferometers we considered the aLIGO sensitivity and the proposed third generation ET. For the space planned interferometers, we considered LISA which is optimized to detect GWs with frequencies of the order of milihertz, gLISA whose concept is very similar to that of LISA, but now the constellation of three spacecrafts is in a geosynchronous orbit, and finally the DECIGO projected sensitivity curve.

Figure \ref{Sh1} shows the predicted PGWs spectral density $S_h(f)$ considering $n$ fixed and varying $\alpha_{M0}$ between positive and negative values according to the stability conditions of the theory. The amplitude of GWs decays after the tensor-modes entry into the horizon, while before the entrance to the horizon the amplitude is practically constant. The time of horizon entry depends on GW frequency along the cosmic expansion. Basically, it is described by the transfer function \cite{TF01}. It is worth mentioning that the resulting spectral density has an oscillatory behavior which is not shown in Figures \ref{Sh1} and \ref{Sh2}. We are showing only the maximum value of $S_h(f)$ which is the relevant quantity for the PGW detection. Also, we take its values divided by a factor of 1/2 due the rapid oscillatory behavior of the spectrum in a detection frequency.


To quantify how much the amplitude changes with respect to GR, we evaluate the quantity $r_h = S_{h, MG}/S_{h, GR}$ numerically which is a constant over the entire frequency range of interest. For the predictions shown in Figure \ref{Sh1}, we find the values displayed in Table \ref{table r}. Therefore, as expected, for $\alpha_{M0} < 0$ we have $S_{h, MG} > S_{h, GR}$, and otherwise for $\alpha_{M0} > 0$. Thus, for $\alpha_{M0} < 0$ we find that the amplitude can change from 0.5\% to 64.9\% , between the assumed values. For $\alpha_{M0} > 0$, we find a decrease in the amplitude ranging from 6.45\% to 99.88\%. Also interesting to notice is that the GW spectral density crosses the DECIGO sensitivity curve. For the other sensitivity curves, the GW spectral density is significantly below of the sensitivity predicted for the experiments.


\begin{table}[] 
\begin{tabular}{l|ccc|ccc} 
\hline
$n$              &          &  2   &       &  & $\Omega_{m0}/2$&     \\ \hline
 $\alpha_{M0}$ & -1.0     & -0.1 & -0.01 & 0.01   & 0.1    & 1.0  \\ \hline
 $r_h$          &  1.649 &  1.051 & 1.005 & 0.936 & 0.513 & 0.001 \\ \hline
\end{tabular}
\caption{Ratio between the spectral density of PGWs in Horndeski gravity and in GR for the parameters of Figure \ref{Sh1}}. 
\label{table r}
\end{table}


Figure \ref{Sh2} shows the predicted PGWs spectral density $S_h(f)$ considering $\alpha_{M0}$ fixed and varying $n$ between values according to the stability conditions of the theory. In this case, the values obtained for $r_h$ are shown in Table \ref{table r2}. Again, note that the predicted spectra cross only the DECIGO sensitivity curve. Figure \ref{DeltaSh} shows $r_h$ as a function of $n$, for the values in which the model is stable.


\begin{table}[] 
\begin{tabular}{l|ccc|ccc} 
\hline
 $\alpha_{M0}$&               &    0.1      &                 &        &  -0.1  &       \\ \hline
 $n$          & $\Omega_{m0}$/2 & $\Omega_{m0}$ & 3$\Omega_{m0}$/2  & 1.5    & 2.5    & 5.0   \\ \hline
 $r_h$       &  0.513        & 0.717       & 0.801           &  1.069 &  1.041 & 1.020 \\ \hline
\end{tabular}
\caption{Ratio between the spectral density of PGWs in Horndeski gravity and in GR for the spectra shown in Figure \ref{Sh2}}. 
\label{table r2}
\end{table}

It is worth stressing that we are assuming a minimum and conservative inflationary model. Once that the PGW amplitude is also very sensitive to the inflationary model and its corresponding parameters, the  spectrum could well be detected by LISA for some of these models (see, e.g., \cite{LISA_inflation} for some results), although one has to pay attention to the observational limits imposed by the parameter $r$.    

\begin{figure}
\includegraphics[width=3.25in,height=2.5in]{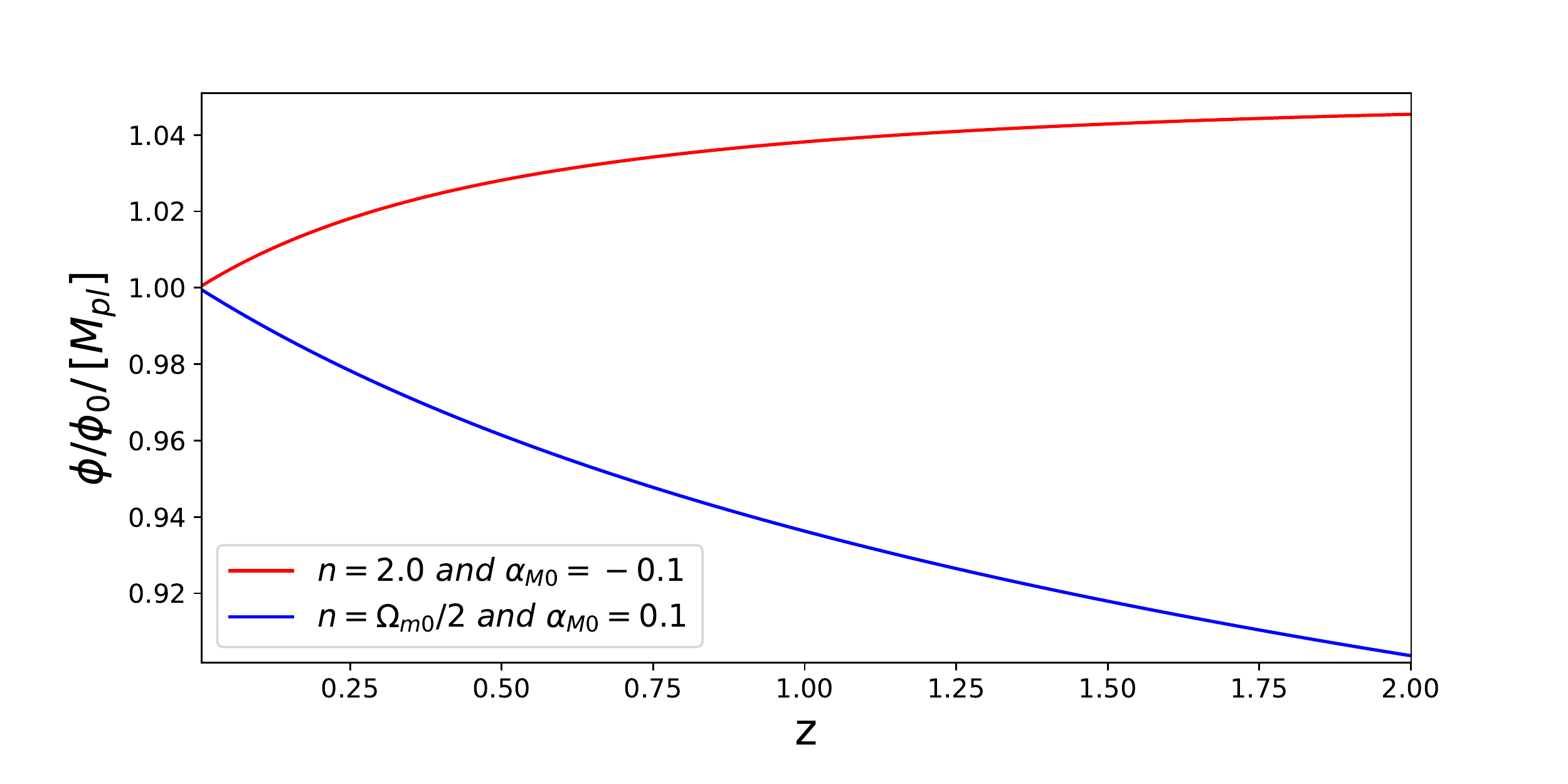} 
\caption{Evolution of $\phi/\phi(z=0)$ at late times. We have considered values of $\alpha_{M0}$ and $n$ within values compatible with the stability of the theory.}
\label{phi}
\end{figure}

Finally, let us briefly discuss our results regarding the construction of viable gravity models. In light of the recent observational bound on $c_T$ from the event GW170817, within the framework of Horndeski gravity, the only option to suppress the terms leading to an anomalous speed is to consider that $G_{4,X} \approx 0$ and $G_5 \approx constant$. Based on these conditions, we can write the running of the Planck mass given by Eq. (\ref{alphaM}) as

\begin{equation}\label{running planck mass}
\alpha_M = \frac{\dot{\phi}}{H} \frac{G_{4, \phi}}{G_4}.
\end{equation}

Based on these considerations, one of the surviving classes of models are the non-minimally coupled theories in which the scalar field $\phi$ has a coupling with the curvature scalar $R$ in the form $G_4(\phi) R$. This class includes the metric $f(R)$ gravity and the Brans-Dicke (BD) theory. For a recent review and classification of models based on the GW170817 event, see \cite{Kase}. The original BD theory, for instance, is obtained by setting $G_4 = \phi$. By substituting this in  Eq. (\ref{running planck mass}) it is possible to obtain $\phi$ as a function of the redshift. The result is shown in Figure \ref{phi}, where it is considered only values of $\alpha_{M0}$ and $n$ compatible with the stability of the theory.

\section{Final remarks}

\label{Conclusions}

We have investigated the PGWs propagation in the context of the Horndeski theories. Assuming a generic parametrization for the $\alpha_i$ functions, the effects are quantified by changes on the PGW spectra.

As a general conclusion, we found that the PGW spectrum is considerably sensitive to the value of $\alpha_{M0}$ such that if $\alpha_{M0} < 0 $ ($\alpha_{M0} > 0 $) the amplitude is larger (smaller) than that predicted by the GR theory. For positive $\alpha_{M0}$, the spectrum is also sensitive to $n$,  but if $\alpha_{M0}$ is negative, only tiny modifications are induced in the amplitude due to the choice of different values of $n$.

The predicted present day spectra were compared with different GW experiments, showing that it can be detected only by the DECIGO detector, at least in the conservative scenario we have adopted. Hence, we found that interesting constraints can be imposed on the parameters of the  Horndeski gravity by combining the future detection of the PGW spectrum with the bounds on the speed of GWs. Therefore, such constraints will be a valuable tool to identify the surviving classes of theories of gravity.

\begin{acknowledgments}
\noindent 

The authors thank the referee for his/her valuable comments and suggestions.  
MESA and JCNA would like to thank the Brazilian agency FAPESP for financial support under the thematic project \# 2013/26258-4. JCNA would like to thank CNPq for partial financial support under grant \# 307217/2016-7

\end{acknowledgments}


\begin{thebibliography}{99}

\bibitem{ligo01} B. P. Abbott et al., Phys. Rev. Lett. {\bf  116}, 061102 (2016), arXiv:1602.03837 [gr-qc].

\bibitem{Gw02}  B. P. Abbott et al., Phys. Rev. Lett. \textbf{116}, 241103 (2016), arXiv:1606.04855 [gr-qc].

\bibitem{Gw03}  B. P. Abbott et al., Phys. Rev. Lett. \textbf{118},  221101 (2017), arXiv:1706.01812 [gr-qc].

\bibitem{Gw04} B. P. Abbott et al., Astrophys. J. \textbf{851}, L35 (2017), arXiv:1711.05578 [astro-ph.HE].

\bibitem{Gw05} B. P. Abbott et al., Phys. Rev. Lett. \textbf{119}, 141101 (2017), arXiv:1709.09660 [gr-qc].

\bibitem{Gw06} B. P. Abbott et al., Astrophys. J. \textbf{832}, L21 (2016), arXiv:1607.07456 [astro-ph.HE].

\bibitem{Gw07} B. P. Abbott et al., Phys. Rev. Lett. \textbf{119}, 161101 (2017), arXiv:1710.05832 [gr-qc].

\bibitem{Gw08}  B. P.  Abbott  et  al., The  Astrophysical  Journal  Letters \textbf{848},  L13  (2017), arXiv:1710.05834.


\bibitem{GW_MG01} T. Baker, E. Bellini, P. G. Ferreira, M. Lagos, J. Noller, and I. Sawicki, Phys. Rev. Lett. \textbf{119}, 251301 (2017), arXiv:1710.06394 [astro-ph.CO].

\bibitem{GW_MG02} P. Creminelli and F. Vernizzi, Phys. Rev. Lett. \textbf{119}, 251302 (2017),  arXiv:1710.05877 [astro-ph.CO].

\bibitem{GW_MG03} J. Sakstein and J. Jain, Phys. Rev. Lett. 119, 251303 (2017),	arXiv:1710.05893 [astro-ph.CO].

\bibitem{GW_MG04} J. M. Ezquiaga and M. Zumalacárregui, Phys. Rev. Lett. \textbf{119}, 251304 (2017),  arXiv:1710.05901 [astro-ph.CO].


\bibitem{GW_MG06} L. Amendola, M. Kunz, I. D. Saltas , and I. Sawicki, Phys. Rev. Lett. \textbf{120}, 131101 (2018), arXiv:1711.04825 [astro-ph.CO].

\bibitem{GW_MG07} J. M. Ezquiaga and M. Zumalacárregui, Front. Astron. Space Sci. \textbf{5} 44 (2018), arXiv:1807.09241 [astro-ph.CO].

\bibitem{Caprini}  C. Caprini and D. G. Figueroa, Class. Quant. Grav. \textbf{35} 16 163001 (2018), arXiv:1801.04268 [astro-ph.CO].

\bibitem{Akrami:2018odb} Y.~Akrami {\it et al.,} [Planck Collaboration], arXiv:1807.06211 [astro-ph.CO].

\bibitem{bicep2} P. A. R. Ade {\it et al.,} [Keck Array/BICEP2 Collaborations], Phys. Rev. Lett. 121, 221301 (2018), arXiv:1810.05216 [astro-ph.CO].

\bibitem{Amaro2017}  P. Amaro-Seoane {\it et al.,} arXiv:1702.00786.
  
\bibitem{Seto2001} N. Seto, S. Kawamura, and T. Nakamura, Phys. Rev. Lett. \textbf{87}, 221103 (2001), arXiv: 0108011 [astro-ph].  

\bibitem{Tinto2016} M. Tinto and J.C.N. de Araujo, Phys.Rev. D \textbf{94}, 8 081101 (2016), arXiv:1608.04790 [astro-ph.IM].

\bibitem{Verbiest2016} J.P.W. Verbiest et al., MNRAS \textbf{458}, 1267 (2016), arXiv:1602.03640 [astro-ph.IM].

\bibitem{MG_GW01}  W. Lin and M. Ishak, Phys. Rev. D \textbf{94}, 123011 (2016), arXiv:1605.03504 [astro-ph.CO].

\bibitem{MG_GW02} A. Nishizawa, Phys. Rev. D \textbf{97}, 104037 (2018), arXiv:1710.04825 [gr-qc].

\bibitem{MG_GW03} S. Arai and A. Nishizawa,
Phys. Rev. D \textbf{97}, 104038 (2018), arXiv:1711.03776 [gr-qc]

\bibitem{MG_GW04} E. Belgacem, Y. Dirian, S. Foffa and M. Maggiore,  Phys. Rev. D \textbf{98}, 023510 (2018),  arXiv:1805.08731 [gr-qc].

\bibitem{MG_GW05} M. E. S. Alves, P. H. R. S. Moraes, J. C. N. Araujo, and M. Malheiro, Phys. Rev. D \textbf{94}, 024032 (2016), arXiv:1604.03874 [gr-qc].

\bibitem{MG_GW06} R. C. Nunes, S. Pan, and E. N. Saridakis, Phys. Rev. D \textbf{98}, 104055 (2018), arXiv:1810.03942 [gr-qc].

\bibitem{MG_GW07} A. Casalino, M. Rinaldi, L. Sebastiani, and S. Vagnozzi, Class. Quant. Grav. \textbf{36} 017001 (2019), arXiv:1811.06830 [gr-qc].

\bibitem{MG_GW08} L. Visinelli, N. Bolis, and S. Vagnozzi, Phys. Rev. D \textbf{97}, 064039 (2018), arXiv:1711.06628 [gr-qc].

\bibitem{MG_GW09} T. Fujita, S. Kuroyanagi, S. Mizuno, S. Mukohyama,  Phys. Lett. B \textbf{7} 89 (2019), arXiv:1808.02381 [gr-qc].

\bibitem{MG_GW10} L. Lombriser and A. Taylor, JCAP \textbf{03} 031 (2016), arXiv:1509.08458 [astro-ph.CO].

\bibitem{MG_GW11} J. B. Jimenez and L. Heisenberg, JCAP \textbf{09} 1809 035 (2018), arXiv:1806.01753 [gr-qc]


\bibitem{MG_GW13}  M. Isi and L. C. Stein, Phys. Rev. D \textbf{98}, 104025 (2018), arXiv:1807.02123 [gr-qc].

\bibitem{MG_GW14} N. J. Cornish, L. O. Beirne, S. R. Taylor, and N. Yunes,  	Phys. Rev. Lett. \textbf{120}, 181101 (2018),  	arXiv:1712.07132 [gr-qc].

\bibitem{MG_GW15} M. Scomparin and S. Vazzoler, arXiv:1903.01502 [gr-qc].

\bibitem{inflation_GW01} S. Kuroyanagi, T. Chiba and N. Sugiyama, Phys. Rev.D \textbf{79} 103501 (2009), arXiv:0804.3249 [astro-ph].

\bibitem{inflation_GW02} K. Saikawa and S. Shirai, JCAP \textbf{05} 035 (2018), arXiv:1803.01038 [hep-ph]. 

\bibitem{inflation_GW03} S. Kuroyanagi, T. Takahashi and S. Yokoyama, JCAP \textbf{02} 003 (2015),  arXiv:1407.4785 [astro-ph.CO].

\bibitem{inflation_GW04} G. Franciolini, G.F. Giudice, D. Racco, A. Riotto, arXiv:1811.08118 [hep-ph].

\bibitem{inflation_GW05} K. Saikawa and S. Shirai, JCAP \textbf{05} 035 (2018), arXiv:1803.01038 [hep-ph].

\bibitem{Deffayet} C. Deffayet, X. Gao, D. A. Steer, and G. Zahariade, Phys. Rev. D \textbf{84}, 064039 (2011), 1103.3260.

\bibitem{Kobayashi} T. Kobayashi, M. Yamaguchi and J. Yokoyama, Prog. Theor. Phys. \textbf{126}, 511 (2011), arXiv:1105.5723 [hep-th].

\bibitem{Horndeski} G. W. Horndeski, Int. J. Theor. Phys. \textbf{10}, 363 (1974).

\bibitem{Horndeski_constraints_01} J. Espejo, S. Peirone, M. Raveri, K. Koyama, L. Pogosian, and A. Silvestri, arXiv:1809.01121 [astro-ph.CO].

\bibitem{Horndeski_constraints_02} E. Bellini, A. J. Cuesta, R. Jimenez, and L. Verde, JCAP \textbf{1602} 02 053 (2016), arXiv:1509.07816 [astro-ph.CO].

\bibitem{Horndeski_constraints_03} L. Perenon, C. Marinoni, and F. Piazza, JCAP \textbf{1701} 01 035 (2017), arXiv:1609.09197 [astro-ph.CO].

\bibitem{Horndeski_constraints_04} D. Alonso, E. Bellini, P. G. Ferreira, and M. Zumalacarregui, Phys. Rev. D \textbf{95}, 063502 (2017), arXiv:1610.09290 [astro-ph.CO].

\bibitem{Horndeski_constraints_05} N. Frusciante, S. Peirone, S. Casas, and N. A. Lima, arXiv:1810.10521 [astro-ph.CO].

\bibitem{Horndeski_constraints_06} C. D. Kreisch and E. Komatsu,  arXiv:1712.02710 [astro-ph.CO].

\bibitem{Horndeski_constraints_07} R. A. Battye, F. Pace, and D. Trinh, Phys. Rev. D \textbf{98}, 023504 (2018), arXiv:1802.09447 [astro-ph.CO].

\bibitem{Horndeski_constraints_08} J. Kennedy, L. Lombriser and A. Taylor, Phys. Rev. D \textbf{98}, 044051 (2018), arXiv:1804.04582 [astro-ph.CO].

\bibitem{Horndeski_constraints_09} M. Brush, E. V. Linder, M. Zumalacarregui, arXiv:1810.12337 [astro-ph.CO].

\bibitem{Horndeski_constraints_10} J. Noller and A. Nicola, arXiv:1811.03082 [astro-ph.CO].


\bibitem{aLIGO} The LIGO Scientific Collaboration, J. Aasi et al., Classical and Quantum Gravity {\bf 32}, 074001 (2015)

\bibitem{DECIGO} S. Kawamura, T. Nakamura, M. Ando et al., Classical and Quantum Gravity {\bf 23}, S125 (2006)

\bibitem{ET} M. Punturo, M. Abernathy, F. Acernese et al, Classical and Quantum Gravity {\bf 27}, 194002  (2010)
 
\bibitem{eLISA} P. Amaro-Seoane, H. Audley, S. Babak et al., arXiv: 1702.00786

\bibitem{Tsujikawa} S. Tsujikawa,  	Lect. Notes Phys. \textbf{892} 97-136 (2015), arXiv:1404.2684 [gr-qc].

\bibitem{Saltas} I. Saltas,  I. Sawicki,  L. Amendola,  and M. Kunz, Phys.
Rev. Lett. \textbf{113} 191101 (2014).

\bibitem{TF01} Y. Watanabe and E. Komatsu, Phys. Rev. D \textbf{73}, 123515 (2006), [astro-ph/0604176].

\bibitem{TF02} L. A. Boyle and P. J. Steinhardt, Phys. Rev. D \textbf{77}, 063504 (2008), [astro-ph/0512014].

\bibitem{TF03}  M. S. Turner, M. J. White and J. E. Lidsey, Phys. Rev. D \textbf{48}, 4613 (1993), [astro-ph/9306029].

\bibitem{Bellini} E. Bellini and I. Sawicki, JCAP \textbf{1407}, 050 (2014),
1404.3713.

\bibitem{alphai_01} J. Kennedy, L. Lombriser and Andy Taylor, Phys. Rev. D \textbf{98}, 044051 (2018), arXiv:1804.04582 [astro-ph.CO].

\bibitem{alphai_02} M. Denissenya and E. V. Linder, arXiv:1808.00013 [astro-ph.CO].

\bibitem{alphai_03} M. Zumalacarregui, E. Bellini, I. Sawicki, J. Lesgourgues, P. G. Ferreira, JCAP \textbf{08} 019 (2017), arXiv:1605.06102 [astro-ph.CO].

\bibitem{Gleyzes2017} J. Gleyzes, Phys. Rev. D \textbf{96}, 063516 (2017), arXiv:1705.04714 [astro-ph.CO]


\bibitem{LISA_inflation} N. Bartolo {\it et al.,} JCAP \textbf{12} 026 (2016), arXiv:1610.06481 [astro-ph.CO].


\bibitem{Kase} R. Kase and S. Tsujikawa, arXiv:1809.08735 [gr-qc].



\end{thebibliography}
\end{document}